\documentclass[12pt]{article}
\usepackage{jhep-mod}
\usepackage{bm}
\usepackage{amssymb,amsmath,amsthm}
\usepackage{mathrsfs}
\usepackage[utf8]{inputenc}
\usepackage{enumerate}
\usepackage{appendix}
\usepackage{graphicx}
\usepackage{dcolumn} 
\usepackage{bm}
\usepackage{multirow}
\usepackage{float}
\usepackage{tikz}
\usepackage{setspace}
\definecolor{purple}{rgb}{1,0,1}
\definecolor{lime}{HTML}{A6CE39} % needs xcolor
\definecolor{lime}{HTML}{A6CE39}
\newcommand{\orcidicon}{%
	\begin{tikzpicture}
	\draw[lime, fill=lime] (0,0) 
		circle [radius=0.16] 
		node[white] {{\fontfamily{qag}\selectfont \tiny ID}};
	\draw[white, fill=white] (-0.0625,0.095) 
		circle [radius=0.007];
	\end{tikzpicture}
	\hspace{-5mm}
}
\newcommand\orcidAlex{{\href{https://orcid.org/0000-0002-1763-3563}{\orcidicon}}}
\newcommand\orcidMatt{{\href{https://orcid.org/0000-0003-1088-6485}{\orcidicon}}}
\begin{document}
%========================================================
\title{\huge{Regular black holes with asymptotically Minkowski cores}}
%========================================================
\author{\Large Alex Simpson\orcidAlex{}{\sf and} Matt Visser\orcidMatt{} }
%========================================================
\affiliation{School of Mathematics and Statistics, Victoria University of Wellington, \\ \null\quad PO Box 600, Wellington 6140, New Zealand}
%========================================================
\emailAdd{alex.simpson@sms.vuw.ac.nz, matt.visser@sms.vuw.ac.nz}
%========================================================
\abstract{
\parindent0pt
\parskip7pt

Standard models of ``regular black holes''  typically have asymptotically de~Sitter regions at their cores. 
Herein we shall consider novel ``hollow'' regular black holes, those  with asymptotically Minkowski cores. 
The reason for doing so is twofold: First, these models greatly simplify the physics in the deep core, and second, one can trade off rather messy cubic and quartic polynomial equations for somewhat more elegant special functions such as exponentials and the increasingly important Lambert $W$ function. While these ``hollow'' regular black holes share many features with the Bardeen/Hayward/Frolov regular black holes there are also significant differences.

\medskip
{\sc Date:} 4 November 2019; 18 November 2019; \LaTeX-ed \today

exponential mass suppression.
}
%=====================================================
\maketitle
%=====================================================
\def\d{{\mathrm{d}}}
\def\tr{{\mathrm{tr}}}
\parindent0pt
\parskip7pt

\vspace{-20pt}
%=====================================================
\section{Introduction}
%====================================================

\enlargethispage{20pt}
It is well established that all static spherically symmetric spacetimes have a line element which can without loss of generality be represented in the following form~\cite{Lorentzian,dirty}:

\begin{equation}\label{Eq1}
    ds^2 = -e^{-2\Phi(r)}\left(1-\frac{2m(r)}{r}\right)dt^2 + \frac{dr^2}{1-\frac{2m(r)}{r}} + r^2\,d\Omega^2_{2},
\end{equation}
Here $\Phi(r)$ and  $m(r)$ are \emph{a priori} arbitrary functions of $r$. 
Historically, most of the prominent models for regular black holes are of this form,
and exhibit an asymptotically de Sitter core with finite central energy density and an equal-but-opposite central pressure. 
(See for instance the  Bardeen/Hayward/Frolov, \emph{etc.}~\cite{Bardeen:1968, Hayward:2005, Frolov:2014, Frolov:2017, viability-regular}) regular black holes, but also the Mazur--Mottola gravastars~\cite{Mazur:2001,Mazur:2004,Visser:2003,Cattoen:2005,
Lobo:2005,Chirenti:2007,MartinMoruno:2011,Lobo:2015,Chirenti:2016},  and Dymnikova's models~\cite{Dymnikova:1999,Dymnikova:2000,Dymnikova:2003,
Dymnikova:2004,Dymnikova:2010}.) 

Presented herein is a new and rather different form of the metric tensor, which is still a regular black hole in the sense of Bardeen~\cite{Bardeen:1968}, but  with an asymptotically Minkowski core (the energy density and associated pressures asymptote to zero). We may make this physical difference mathematically explicit by examining the required conditions on the two functions  $\Phi(r)$ and $m(r)$ appearing in the metric~(\ref{Eq1}):

\begin{itemize}
    \item \emph{de Sitter core:} We would require that $\lim_{r\rightarrow 0}\rho(r) = \rho_0 \neq 0$; that is $\rho(r) = O(1)$. For static spherically symmetric spacetimes we have $\rho(r)=m'(r)/4\pi r^{2}$~\cite{Lorentzian,Wald,MTW,Hell}, so this implies $m(r) = O(r^3)$. For $\Phi(r)$ it is sufficient to demand $\Phi(r) = O(1)$. 
    \item \emph{Minkowski core:} We would require that $\lim_{r\rightarrow 0}\rho(r) = 0$; that is  $\rho(r) = o(r)$. This in turn implies $m(r) = o(r^3)$. For $\Phi(r)$ it is once again sufficient to demand $\Phi(r)=O(1)$.
\end{itemize}
For simplicity/tractability we therefore might as well immediately set $\Phi(r) = 0$, which will preserve all of the key features our desired physics. 
($\Phi(r)=0$ spacetimes have a long and respected history. See specifically references~\cite{Lorentzian} and~\cite{Jacobson:2007, Kiselev:1,Kiselev:2}, and more generally references~\cite{Dymnikova:1999,Dymnikova:2000,Dymnikova:2003,
Dymnikova:2004,Dymnikova:2010}.) 
The choice for $m(r)$ is more subtle. Presented herein is a specific choice of mass function which has the effect of ``exponentially suppressing'' the mass of the centralised object as one nears the coordinate location $r = 0$; guaranteeing that the object has an asymptotically Minkowski core. That is to say we analyse the specific line element:

\begin{equation}
    ds^2 = -\left(1-\frac{2m\,e^{-a/r}}{r}\right)dt^2 + \frac{dr^2}{\left(1-\frac{2m\,e^{-a/r}}{r}\right)} + r^2 \; d\Omega^2_{2} .
\end{equation}

Conducting a standard general relativistic analysis of the resulting  spacetime, we shall demonstrate that this metric does indeed correspond to a regular black hole in the sense of Bardeen~\cite{Bardeen:1968}. We find that in this toy model the resulting curvature tensors and curvature invariants are significantly simpler than for Bardeen/Hayward/Frolov, at the cost of having many physically interesting  features defined by the Lambert $W$ function, one of the special functions of mathematics~\cite{Valluri:2000, Valluri:2009, Boonserm:2008, Boonserm:2010,Boonserm:2013, Boonserm:2018exp, Sonoda:2013a, Sonoda:2013b, Corless:1996, Vial:2012, Stewart:2011, Stewart:2012, Visser:2018-LW}.

\enlargethispage{30pt}
A rather different (extremal) version of this model spacetime, based on nonlinear electrodynamics, has been previously discussed by Culetu~\cite{Culetu:2013}, with follow-up on some aspects of the non-extremal case in references~\cite{Culetu:2014, Culetu:2015a, Culetu:2015b}. See also~\cite{Junior:2015,Rodrigues:2015}. 
Part of the GRF essay~\cite{Xiang:2013} is based on a mass function of the form $m(r)= m\,e^{-a^2/r^2}$. 
Another very different type of exponential modification of Schwarzschild and Kerr geometries, obtained by inserting factors of the form $1-a\exp(-b^3/r^3)$ into the spacetime metric, has been discussed by Takeuchi~\cite{Takeuchi:2016}.
The overall framework these authors are working with differs significantly from our own.

%=====================================================
\section{Metric analysis}\label{sec:metric}
%=====================================================

First let us note that the metric is static and spherically symmetric. The areas of surfaces of spherical symmetry of constant $r$-coordinate are trivial, given by the area function $A(r)=4\pi r^{2}$, which we can clearly see is minimised at $r=0$.

Note that this representation of the metric corresponds to the central mass being $r$-dependent in the following manner: $m(r)=m\,e^{-a/r}$. We immediately have $a\in\mathbb{R}^{+}$ in order to ensure the mass is being exponentially ``suppressed'' as $\vert r\vert\rightarrow 0$; in contrast if $a=0$ we simply have the Schwarzschild solution, and if $a<0$ we have an altogether different scenario where asymptotic behaviour for small $r$ indicates massive exponential ``growth''. For $a>0$ the exponential expression factor has the  properties:
\begin{equation}
    \lim_{r\rightarrow 0^{+}}e^{{-a/r}}=0 , \qquad \lim_{r\rightarrow 0^{-}}e^{{-a/r}}=+\infty .
\end{equation}
The metric is $C^\infty$ smooth but not $C^\omega$ analytic at coordinate location $r=0$. Looking at the behaviour as $r\rightarrow 0^{-}$ we see that the region $r<0$ can safely be omitted from the analysis; the severe discontinuity at $r=0$ implies that behaviour in the negative $r$ domain is grossly unphysical. This does not affect our coordinate patch, as $r=0$ also marks a coordinate singularity. (Demonstration that this is a coordinate singularity and not a curvature singularity is a corollary of analysis in \S\ref{suppressedtensors}.) We may trivially avoid these ``issues'' by enforcing $r\in\mathbb{R}^{+}$, \emph{i.e.} strictly remaining within ``our'' universe, which is the primary region of interest. In view of the diagonal metric environment, we may now examine horizon locations for the spacetime by setting $g_{tt}=0$:
\begin{equation}
    g_{tt} = 0 \qquad \Longrightarrow \quad 
    r = -\frac{a}{W\!\!\left(-\frac{a}{2m}\right)} \qquad \Longrightarrow \quad 
    r = 2m \ e^{W\!\left(-\frac{a}{2m}\right)} .
\end{equation}

We have a coordinate location for the horizon defined \emph{explicitly} in terms of the Lambert $W$ function. Given that we have taken $a>0$ and $r>0$, we must enforce the output of the Lambert $W$ function to be negative. This presents two possibilities:
\begin{itemize}
\item 
Taking the $W_{0}\left(x\right)$ branch of the real-valued Lambert $W$ function:
\begin{equation}
 W_{0}\!\left(-\frac{a}{2m}\right)<0 \quad \Longrightarrow \quad a\in \ \left(0, \ \frac{2m}{e}\right] .
\end{equation}
Provided $a$ lies in this interval we have a well-defined coordinate location for a horizon when taking the $W_{0}(x)$ branch of the Lambert $W$ function. Keeping in mind that fixing $a$ in this interval causes $W_{0}\left(-\frac{a}{2m}\right)\in\left[-1, \ 0\right)$, the possible coordinate locations of \emph{this} horizon are given by $r_{H}\in \ [a, +\infty)$.
    
\item 
Taking the $W_{-1}\left(x\right)$ branch of the real-valued Lambert $W$ function:
    
The $W_{-1}(x)$ branch only returns outputs for $x\in \ \left[-\frac{1}{e}, 0\right)$, hence we have the same restriction on $a$ as before to ensure a defined coordinate location for the horizon; that $a\in \ \left(0, \frac{2m}{e}\right]$. The range of the $W_{-1}(x)$ branch is entirely negative so all possible solutions will correspond to $r_H >0$. However the difference is that fixing $a$ in this interval causes $W_{-1}\left(-\frac{a}{2m}\right)\in \ [-1, -\infty)$, hence the possible coordinate locations for \emph{this} horizon are given by $r_{H}\in \ (0, a]$.
\end{itemize}
In view of the possible horizon locations corresponding to the two possible branches, we see that there is an \emph{outer horizon} at coordinate location $r = 2m \ e^{W_{0}\left(-\frac{a}{2m}\right)}$, and an \emph{inner horizon} at $r = 2m \ e^{W_{-1}\left(-\frac{a}{2m}\right)}$. 
When $a=\frac{2m}{e}$, we see that $W_{-1}\left(-\frac{1}{e}\right) = W_{0}\left(-\frac{1}{e}\right) = -1$; the two horizons merge and we in fact have an extremal black hole. 

It follows that in order to have $r_H>0$, we require $a\in \left(0, \frac{2m}{e}\right]$. Then subject to the choice of branch for the Lambert $W$ function, we find an inner horizon located in the region $r_{H}\in \ (0, a)$, and an outer horizon in the region $r_{H}\in (a, +\infty)$. In both cases the geometry is certainly modelling a black hole region of some description; it remains to demonstrate that the spacetime is gravitationally nonsingular in order to show this is a regular black hole in the sense of Bardeen~\cite{Bardeen:1968}. 

As an aside let us take a look at what happens to the geometry when $a > \frac{2m}{e}$:
\begin{eqnarray}
a > \frac{2m}{e} \qquad &\Longrightarrow& \qquad
 -\frac{a}{2m} < -\frac{1}{e}    \qquad \Longrightarrow\qquad
  W\left(-\frac{a}{2m}\right) \ \mbox{is undefined}.
\end{eqnarray}
Therefore there are no horizons in this geometry. It follows that when $a > \frac{2m}{e}$, there is no black hole of any kind; the geometry is modelling something qualitatively different. 

In any of these situations the geometry admits an almost-global coordinate patch with the following domains:
$t\in\left(-\infty, +\infty\right)$, $r\in\mathbb{R}^{+}$, $\theta\in(0,\pi)$, and $\phi\in(-\pi, \pi)$. We now examine the non-zero components of the curvature tensors, as well as the curvature invariants, to show that, for $a\in\left(0, \frac{2m}{e}\right]$, this metric does indeed model a regular black hole geometry.

%=====================================================
\section{Curvature tensors and curvature invariants}\label{suppressedtensors}
%=====================================================

Before proceeding it is prudent to introduce a relevant piece of mathematical detail: For any polynomial function $p(r)$, we have ${e^{-a/r}}/{p(r)} \rightarrow 0$ as $\vert r\vert\rightarrow 0$. That is, the exponential expression dominates the asymptotic behaviour for small $r$. Keeping this in mind, let us examine the mixed non-zero curvature tensor components and the curvature invariants.

\noindent
Ricci scalar:
\begin{equation}
    R = \frac{2m\,a^{2}\,e^{-a/r}}{r^{5}} .
\end{equation}
Ricci tensor non-zero components:
\begin{equation}
    R^{t}{}_{t} = R^{r}{}_{r} = \frac{m\,a\left(a-2r\right)e^{-a/r}}{r^{5}} , \qquad 
    R^{\theta}{}_{\theta} = R^{\phi}{}_{\phi} = \frac{2m\,a\,e^{-a/r}}{r^{4}} .
\end{equation}
Einstein tensor non-zero components:
\begin{equation}\label{einsteinsuppressed}
    G^{t}{}_{t} = G^{r}{}_{r} = -\frac{2m\,a\,e^{-a/r}}{r^{4}} , \qquad 
    G^{\theta}{}_{\theta} = G^{\phi}{}_{\phi} = -\frac{ma\left(a-2r\right)e^{-a/r}}{r^{5}} .
\end{equation}
Riemann tensor non-zero components:
\begin{eqnarray}
    R^{tr}{}_{tr} &=& \frac{m\left(a^{2}-4ar+2r^{2}\right)e^{-a/r}}{r^{5}} , \nonumber \\
    && \nonumber \\
    R^{t\theta}{}_{t\theta} &=& R^{t\phi}{}_{t\phi} = R^{r\theta}{}_{r\theta} = R^{r\phi}{}_{r\phi} = \frac{m\left(a-r\right)e^{-a/r}}{r^{4}} , \nonumber \\
    && \nonumber \\
    R^{\theta\phi}{}_{\theta\phi} &=& \frac{2m\,e^{-a/r}}{r^{3}} .
\end{eqnarray}
Weyl tensor non-zero components:
\begin{eqnarray}
    -\frac{1}{2}C^{tr}{}_{tr} = -\frac{1}{2}C^{\theta\phi}{}_{\theta\phi} = C^{t\theta}{}_{t\theta} &=& C^{t\phi}{}_{t\phi} = C^{r\theta}{}_{r\theta} = C^{r\phi}{}_{r\phi}  \nonumber \\
    &=& -\frac{m\left(a^{2}-6ar+6r^{2}\right)e^{-a/r}}{6r^{5}} .
\end{eqnarray}
The Ricci contraction $R_{ab}\,R^{ab}$:
\begin{equation}
    R_{ab}\,R^{ab} = \frac{2m^{2}a^{2}\left(a^{2}-4ar+8r^{2}\right)e^{-2a/r}}{r^{10}} .
\end{equation}
The Kretschmann scalar $R_{abcd}\,R^{abcd}$:
\begin{equation}
    R_{abcd}\,R^{abcd} = \frac{4m^{2}\left(a^{4}-8a^{3}r+24a^{2}r^{2}-24ar^{3}+12r^{4}\right)e^{-2a/r}}{r^{10}} .
\end{equation}
\enlargethispage{20pt}
The Weyl contraction $C_{abcd}\,C^{abcd}$:
\begin{equation}
    C_{abcd}\,C^{abcd} = \frac{4m^{2}\left(a^{2}-6ar+6r^{2}\right)^{2}e^{-2a/r}}{3r^{10}} .
\end{equation}

As $\vert r\vert\rightarrow +\infty$ we have $e^{-a/r}\rightarrow 1$, and all non-zero tensor components and invariants become proportional to some Laurent polynomial function of $r$ (\emph{i.e.} all asymptote to $O(r^{-n})$). For large $r$, all non-zero components and invariants tend to zero, consistent with the fact that asymptotic infinity models Minkowski space. As $r\rightarrow 0^{+}$, all non-zero tensor components and invariants also asymptotically head to zero (they become $o(r)$). This is of course represents the fact that the core of the black hole region is asymptotically Minkowski. So we have a geometry which approaches Minkowski both near the centre and at asymptotic infinity, with some maximised region of curvature located in between. We may conclude that all non-zero tensor components and invariants are most certainly globally finite, and as such the geometry possesses no curvature singularities as predicted --- we are indeed dealing with a regular black hole spacetime in the sense of Bardeen.

%=======================================================
\section[Surface gravity, Hawking temperature, and horizon area]
{Surface gravity, Hawking temperature, and horizon area}
\label{sec:3+1}
%========================================================

Let us calculate the surface gravity at the two horizons when $a\in(0,\frac{2m}{e})$. 
The Killing vector which is null at the event horizon is $\xi=\partial_{t}$, that is 
$\xi^\mu=(1,0,0,0)^\mu$. This yields the norm:
\begin{equation}
    \xi^{\mu}\xi_{\mu} = g_{\mu\nu}\xi^{\mu}\xi^{\nu} = g_{tt} = -\left(1-\frac{2m\,e^{-\frac{a}{r}}}{r}\right) .
\end{equation}

Then we have the following relation for the surface gravity $\kappa$ (see for instance~\cite{Wald, MTW, Hell}):
\begin{equation}
    \nabla_{\nu}\left(-\xi^{\mu}\xi_{\mu}\right) = 2\kappa\xi_{\nu}.
\end{equation}
That is:
\begin{equation}
 \nabla_{\nu}\left(1-\frac{2me^{-\frac{a}{r}}}{r}\right) = 2\kappa\xi_{\nu} .
\end{equation}
Alternatively we can apply the formalism of reference~\cite{dirty}. 

\clearpage
Now perform a case-by-case analysis for each horizon:
\begin{itemize}
    \item \emph{Outer horizon:} The outer horizon is located at radial coordinate $r = 2m\,e^{W_{0}\left(-\frac{a}{2m}\right)}$; we therefore have:   
\begin{eqnarray}
    \kappa_{outer} &=& \frac{1}{2}\partial_{r}\left(1-\frac{2me^{-\frac{a}{r}}}{r}\right)\Bigg\vert_{r=2m\,e^{W_{0}\left(-\frac{a}{2m}\right)}} 
    \nonumber\\
    &=& \kappa_{Schwarzschild}\left\lbrace-\frac{2mW_{0}\left(-\frac{a}{2m}\right)}{a}\left[1+W_{0}\left(-\frac{a}{2m}\right)\right]\right\rbrace .
\end{eqnarray}
The Hawking temperature for our regular black hole is~\cite{Wald, MTW, Hell}:
\begin{equation}
   T_{outer} = \frac{\hslash\kappa_{outer}}{2\pi k_{B}} = T_{Schwarzschild}\left\lbrace-\frac{2mW_{0}\left(-\frac{a}{2m}\right)}{a}\left[1+W_{0}\left(-\frac{a}{2m}\right)\right]\right\rbrace .
\end{equation}

For the area of the outer horizon we have:
\begin{equation}
    A_{outer} = 4\pi\left(2m\,e^{W_{0}\left(-\frac{a}{2m}\right)}\right)^{2} = \frac{4\pi a^2}{\left[W_{0}\!\left(-\frac{a}{2m}\right)\right]^{2}} .
\end{equation}

As a sanity check, let us check the behaviour of $\kappa_{outer}$ when $a=0$ (corresponding to the Schwarzschild solution). We have: 
\begin{eqnarray}
    \lim_{a\rightarrow 0}\kappa_{outer} &=& \kappa_{Schwarzschild}\;
    \lim_{a\rightarrow 0}\left\lbrace-\frac{2m\,W_{0}\left(-\frac{a}{2m}\right)}{a}\left[1+W_{0}\left(-\frac{a}{2m}\right)\right]\right\rbrace , \nonumber \\
    && \nonumber \\
    &=& \kappa_{Schwarzschild}\;\lim_{a\rightarrow 0}\left\lbrace -\frac{2m\,W_{0}\left(-\frac{a}{2m}\right)}{a}\right\rbrace , \nonumber \\
    && \nonumber \\
    &=& \kappa_{Schwarzschild}\; \lim_{a\rightarrow 0}e^{-W_{0}\left(-\frac{a}{2m}\right)} 
    \nonumber\\
    &=& \kappa_{Schwarzschild}.
\end{eqnarray}
This is the expected result.

\clearpage
\item \emph{Inner horizon:} The inner horizon is located at $r = 2m \ e^{W_{-1}\left(-\frac{a}{2m}\right)}$; and through similar analysis we have the following: 
\begin{eqnarray}
    \kappa_{inner} &=& \kappa_{Schwarzschild}\left\lbrace-\frac{2m\,W_{-1}\!\left(-\frac{a}{2m}\right)}{a}\left[1+W_{-1}\!\left(-\frac{a}{2m}\right)\right]\right\rbrace \ ; \\
    && \nonumber \\
    T_{inner} &=& T_{Schwarzschild}\left\lbrace-\frac{2m\,W_{-1}\!\left(-\frac{a}{2m}\right)}{a}\left[1+W_{-1}\!\left(-\frac{a}{2m}\right)\right]\right\rbrace \ ; \\
    && \nonumber \\
    A_{inner} &=& \frac{4\pi a^{2}}{\left[W_{-1}\!\left(-\frac{a}{2m}\right)\right]^{2}} .
\end{eqnarray}

\item \emph{Extremal horizon:} Extremality arises when the two horizons merge; this occurs when $a=2m/e$,  at coordinate location $r = 2m/e = a$, where $W_0(-{1\over e})=-1=W_{-1}(-{1\over e})$. Consequently $\kappa_{extremal}=0$, while $T_{extremal}=0$, and $A_{extremal} = 4\pi a^2$.

\end{itemize}

%===========================================
\section{Stress-energy tensor and energy conditions}
%===========================================

\enlargethispage{40pt}
Consider the Einstein field equations for this spacetime, defining the stress-energy-momentum tensor by $G^{\mu}{}_{\nu}=8\pi \, T^{\mu}{}_{\nu}$. From equation~(\ref{einsteinsuppressed}) we have:
\begin{equation}\label{stresssuppressed}
    T^{\mu}{}_{\nu} = \begin{bmatrix}
    -\rho & 0 & 0 & 0 \\
    0 & p_{\parallel} & 0 & 0 \\
    0 & 0 & p_{\perp} & 0 \\
    0 & 0 & 0 & p_{\perp}
    \end{bmatrix}.
\end{equation}
Specifically, for the principal pressures:
\begin{eqnarray}
    \rho &=& - p_{\parallel} = \frac{2m\,a\,e^{-a/r}}{8\pi r^{4}} , \nonumber \\
    && \nonumber \\
    p_{\perp} &=& -\frac{m\,a\left(a-2r\right)e^{-a/r}}{8\pi r^{5}} .
\end{eqnarray}
Let us examine where the energy density is maximised for this spacetime (this is of specific interest due to the exponential suppression of the mass -- we wish to see how the suppression affects the distribution of energy densities in through the geometry):
\begin{equation}
    \frac{\partial{\rho}}{\partial{r}} = \frac{2m\,a\,e^{-a/r}}{8\pi r^{6}}\left(a-4r\right) .
\end{equation}
We see that $\rho$ is maximised at coordinate location $r=\frac{a}{4}$. Let us now analyse the various energy conditions~\cite{Barcelo:2002,Friedman:1993,Flanagan:1996,Ford:1994,Hartman:2016,Roman:1986,Buniy:2005,Buniy:2006,Martin-Moruno:2013a,Martin-Moruno:2013b,Martin-Moruno:2015,Martin-Moruno:2017} and see whether they are violated in our spacetime.
%========================================================
\subsection{Null energy condition}
%========================================================
In order to satisfy the null energy condition, we require that both $\rho+p_{\parallel}\geq 0$ \emph{and} $\rho+p_{\perp}\geq 0$ globally in our spacetime. Let us first consider $\rho+p_{\parallel}$ and note that it is identically zero. 
(This result is common to all $\Phi(r)=0$ spacetimes, see~\cite{Lorentzian,Jacobson:2007,Kiselev:1,Kiselev:2}.)
Now noting that for all $\Phi(r)=0$ spacetimes $\rho+p_{\perp} = \frac{r}{2}\,\rho' $ (see reference~\cite{Kiselev:2}) we have:
\begin{equation}
    \rho+p_{\perp} = \frac{r}{2}\,\rho' = \frac{m\,a\,e^{-a/r}}{8\pi r^5}\left(a-4r\right) .
\end{equation}
This changes sign when $r=\frac{a}{4}$. We therefore have the somewhat non-typical situation where the radial NEC is satisfied by the geometry whilst the transverse NEC is violated in the deep core. 

%========================================================
\subsection{Strong energy condition}
%========================================================
In order to satisfy the strong energy condition (SEC), one of the conditions we require is that $\rho+p_{\parallel}+2p_{\perp}\geq 0$ globally in our spacetime. Evaluating:
\begin{equation}
    \rho+p_{\parallel}+2p_{\perp} = 2p_{\perp} = -\frac{ma\left(a-2r\right)e^{-a/r}}{4\pi r^{5}} .
\end{equation}
This is negative for the region $r<\frac{a}{2}$, and we may conclude that the SEC is violated in the deep core --- this is consistent with expectations.

%========================================================
\section{Comparison with existing regular black hole models}
%========================================================

The primary mathematical benefit of this new model is the relative tractability of the curvature tensors and invariants analysis. Let us briefly compare some of the corresponding results for the Bardeen, Hayward, and Frolov regular black holes. The Bardeen regular black hole is defined by the line element~\cite{Bardeen:1968}:
\begin{equation}\label{Bardeen}
    ds^2 = -\left(1-\frac{2mr^2}{\left[r^2+\left(2m\ell^2\right)^{\frac{2}{3}}\right]^{\frac{3}{2}}}\right)dt^2 + \frac{dr^2}{\left(1-\frac{2mr^2}{\left[r^2+\left(2m\ell^2\right)^{\frac{2}{3}}\right]^{\frac{3}{2}}}\right)} + r^2\;d\Omega^2_{2}.
\end{equation}
For the Hayward model we have~\cite{Hayward:2005}:
\begin{equation}
    ds^2 = -\left(1-\frac{2mr^2}{r^3+2m\ell^2}\right)dt^2 + \frac{dr^2}{\left(1-\frac{2mr^2}{r^3+2m\ell^2}\right)} + r^2\; d\Omega^{2}_{2},
\end{equation}
and for the static case of the Frolov regular black hole we have~\cite{Frolov:2017}:
\begin{equation}
    ds^2 = -\left(1-\frac{2mr^2}{r^3+2m\ell^2+\ell^3}\right)dt^2 + \frac{dr^2}{\left(1-\frac{2mr^2}{r^3+2m\ell^2+\ell^3}\right)} + r^2\; d\Omega^{2}_{2}.
\end{equation}
where in all three cases $\ell$ is some length scale, typically identified with the Planck length~\cite{viability-regular}. Now consider a direct comparison of the resulting Einstein tensors:

\begin{itemize}
\item 
\emph{Bardeen:} The Einstein tensor has non-zero components:
    \begin{equation*}
    G^{t}{}_{t} = G^{r}{}_{r} = \frac{-6m\left(2ml^2\right)^{\frac{2}{3}}}{\left(r^2+\left(2ml^2\right)^{\frac{2}{3}}\right)^{\frac{11}{2}}}\Bigg\lbrace r^6+3r^4\left(2ml^2\right)^{\frac{2}{3}}+3r^2\left(2ml^2\right)^{\frac{4}{3}}+\left(2ml^2\right)^{2}\Bigg\rbrace , \nonumber
\end{equation*}
\begin{equation}
    G^{\theta}{}_{\theta} = G^{\phi}{}_{\phi} = \frac{3m\left(2ml^2\right)^{\frac{2}{3}}\left(3r^2-2\left(2ml^2\right)^{\frac{2}{3}}\right)}{\left(r^2+\left(2ml^2\right)^{\frac{2}{3}}\right)^{\frac{7}{2}}} .
 \end{equation}
As $r\to0$ we have $G^a{}_b \to -(3/\ell^2) \,\delta^a{}_b$.     
\item 
\emph{Hayward:} The Einstein tensor has non-zero components:
    \begin{eqnarray}
        G^{t}{}_{t} &=& G^{r}{}_{r} = -\frac{12m^2\ell^2}{\left(2m\ell^2+r^3\right)^{2}} , \nonumber \\
        && \nonumber \\
        G^{\theta}{}_{\theta} &=& G^{\phi}{}_{\phi} = -\frac{24m^2\ell^2\left(m\ell^2-r^3\right)}{\left(2m\ell^2+r^3\right)^{3}} .
    \end{eqnarray}
As $r\to0$ we have $G^a{}_b \to -(3/\ell^2) \,\delta^a{}_b$.       

\item 
\emph{Frolov:} The Einstein tensor has non-zero components:
    \begin{eqnarray}
  G^{t}{}_{t} &=& G^{r}{}_{r} = -\frac{6m\ell^2\left(2m+\ell\right)}{\left(r^3+2m\ell^2+\ell^3\right)^{2}} \ ; \nonumber \\
  && \nonumber \\
G^{\theta}{}_{\theta} &=& G^{\phi}{}_{\phi} = 
\frac{6m\ell^{2}\left(2r^3\left[2m+\ell\right]-\ell^2(2m+\ell^2)^2 \right)}{\left(r^{3}+2m\ell^{2}+\ell^{3}\right)^{3}} \ .
\end{eqnarray}
As $r\to0$ we have $G^a{}_b \to -(6m/[\ell^2(2m+\ell)]) \,\delta^a{}_b$.     

\item \emph{Exponential supression:} As displayed in equation~(\ref{einsteinsuppressed}), we have:
    \begin{equation}
    G^{t}{}_{t} = G^{r}{}_{r} = -\frac{2mae^{-a/r}}{r^{4}} , \quad G^{\theta}{}_{\theta} = G^{\phi}{}_{\phi} = -\frac{ma\left(a-2r\right)e^{-a/r}}{r^{5}} .
    \end{equation}
 As $r\to0$ we have $G^a{}_b \to 0$.  
\end{itemize}

This relative simplicity comes at the cost of representing important physical features  in terms of the Lambert W function. 

%\clearpage
For example we may compare the horizon locations:
\begin{description}
\item[Bardeen:] 
$r_{H} \in \left\lbrace 
\ell+\mathcal{O}(\ell^{5/3}/m^{2/3}), 2m+\mathcal{O}(\ell^{4/3}/m^{1/3})
\right\rbrace$;

\item[Hayward/Frolov:] 
$r_{H} \in \left\lbrace 
\ell+\mathcal{O}(\ell^2/m), 2m+\mathcal{O}(\ell^2/m)
\right\rbrace$;

\item[Exponential suppression:] 
$r_{H} \in \left\lbrace 
2m\,e^{W_{-1}\left(-\frac{a}{2m}\right)}, 2m\,e^{W_{0}\left(-\frac{a}{2m}\right)}
\right\rbrace $.
\end{description}

While this may seem somewhat unwieldy, the authors argue that the Lambert W function is now such a crucial tool in many quite standard GR analyses~\cite{Boonserm:2013,Boonserm:2018exp} that theoreticians would do well to become more comfortable in dealing with its subtleties.

Physically, the crucial comparison is that the canonical regular black hole solutions (Bardeen, Hayward, Frolov, \emph{etc.}) have an asymptotically de Sitter core, whilst the model presented herein has an asymptotically Minkowski core. Whether there are scenarios where this has significant ramifications on black hole evolution or evaporation is a subject for further research.

%========================================================
\section{Generalised models}
%========================================================

Now consider a generalized model of the form 
\begin{equation}\label{generalized}
    ds^2=-\left(1-\frac{2m\exp({{-a^{n}/r^{n}}})}{r}\right)dt^2
    +\frac{dr^2}{\left(1-\frac{2m\exp({{-a^{n}/r^{n}}})}{r}\right)}+r^2\, d\Omega^{2}_{2} \ ; 
    \quad a, n \in\mathbb{R^{+}} .
\end{equation} 
This corresponds to the central mass being $r$-dependent in the following manner: $m(r)=m \exp({-a^{n}/r^{n}})$. We may immediately enforce  $a\in\mathbb{R}^{+}$ and $r\in\mathbb{R}^{+}$. We examine horizon locations by setting $g_{tt}=0$. Thence:
\begin{eqnarray}
    r_H &=& 2m\exp\left({-\frac{a^n}{r_H^n}}\right)
    \quad \Longrightarrow \quad r_H^n = (2m)^{n}\exp\left({-\frac{na^n}{r_H^n}}\right).
\end{eqnarray}
That is:
\begin{equation}
r_H = 2m\;\exp\left({\frac{W\!\left(-\frac{na^{n}}{(2m)^{n}}\right)}{n}}\right) .
\end{equation}
For general values of $n$ this defines a coordinate location of the horizon explicitly in terms of the real-valued branches of the Lambert $W$ function.
While we could repeat the entire analysis for general values of $n$ we feel there is little extra insight to be gained --- all such models will have asymptotically Minkowski cores.

%========================================================
\section{Discussion}\label{sec:dis}
%========================================================

The model spacetime presented above represents a regular black hole geometry in the sense of Bardeen when the parameter $a\in\left(0, \frac{2m}{e}\right]$, and accordingly violates the SEC. The model satisfies the radial NEC, (there is no wormhole throat), but violates the tangential NEC in the deep core when $r<\frac{a}{4}$. The energy density $\rho$ is maximised at $r=\frac{a}{4}$, and the exponential expression present in the metric implies that the regular black hole has an asymptotically Minkowski core as $r\rightarrow 0$. The curvature of the geometry is asymptotically flat at infinity, at the core, and has some maximal peak in between. We can calculate surface gravity at each horizon, and consequently the Hawking temperature. The curvature tensors and invariants are significantly simpler than for canonical regular black hole solutions (Bardeen/Hayward/Frolov \emph{etc.}), at the cost of having some important physical features defined explicitly in terms of the Lambert $W$ function. We may extend the analysis beyond $n=1$ to arbitrary (real) values of $n$ whilst still preserving all desired physics. The authors contend that this regular black hole model is mathematically interesting due to its  tractability, and physically interesting due to the non-standard asymptotically Minkowski core.
%%%%%%%%%%%%%%%%%%%%%%%%%%%%%%%%%%%%%%%%%

%========================================================
\section*{Acknowledgments}
%========================================================
MV acknowledges financial support via the Marsden Fund administered by the Royal Society of New Zealand. AS acknowledges financial support via the PhD Doctoral Scholarship provided by Victoria University of Wellington.

%========================================================
%========================================================

%========================================================
\end{document}